\documentclass[11pt,a4paper]{article}
\usepackage[utf8]{inputenc}
\usepackage{jheppub}
\usepackage[numbers,sort&compress]{natbib}
\usepackage{amsmath} \usepackage{amsfonts} \usepackage{amssymb}
\usepackage{wasysym}
\usepackage[dvipsnames]{xcolor}
\usepackage{graphicx}
\usepackage{multirow}
\usepackage{here}
\usepackage{epsfig}
\usepackage{epstopdf}
\usepackage{soul}
\usepackage{hyperref}
\usepackage{float}

\usepackage{xcolor}
\usepackage{graphicx}
\usepackage{multirow}
\usepackage{ulem}
\usepackage{array}
\usepackage{journals_aastex}

\newcolumntype{L}{>{\displaystyle}l}
\newcolumntype{C}{>{\displaystyle}c}
\newcolumntype{R}{>{\displaystyle}r}

\newcommand{\be}{\begin{equation}}
\newcommand{\ee}{\end{equation}}
\newcommand{\bea}{\begin{eqnarray}}
\newcommand{\eea}{\end{eqnarray}}

\title{The Grand Canonical Multiverse and the Small Cosmological Constant}
\author[a,b,c]{Ido Ben-Dayan}
\author[d]{Merav Hadad}
\author[a]{Amir Michaelis}

\affiliation[a]{Department of Physics, Ariel University, Ariel, POB 3, 4070000, Israel}
\affiliation[b]{Berkeley Center for Cosmological Physics, University of California, Berkeley, CA 94720, USA}
\affiliation[c]{Department of Physics, University of California, Berkeley, CA 94720, USA}
\affiliation[d]{Department of Natural Sciences, The Open University of Israel P.O.B. 808, Raanana 43107, Israel}
\emailAdd{ido.bendayan@gmail.com}
\emailAdd{meravha@openu.ac.il}
\emailAdd{amirmi@ariel.ac.il}
\date{June 2021}

\abstract{
 We consider the Multiverse as an ensemble of universes. Using standard statistical physics analysis we get that the Cosmological Constant (CC) is exponentially small. The small and finite CC is achieved without any anthropic reasoning. We then quantize the CC. The quantization allows a precise summation of the possible contributions and using the measured value of the CC yields a prediction on the temperature of the Multiverse that we define. Furthermore, quantization allows the interpretation of a single Universe as a superposition of different eigenstates with different energy levels rather than the existence of an actual Multiverse.}
\dedicated{
Dedicated to the memory of my father Nissim Ben-Dayan z"l \\
who is still my father in all the universes of the Multiverse - Ido.}
\begin{document}

\maketitle

\section{Introduction}
Slow-roll Inflation is a successful paradigm for describing the Early Universe. 
The paradigm describes the creation and evolution of small quantum fluctuations on top of a classical (slow-roll) evolution of a nearly de-Sitter (dS) phase. The classical evolution solves the flatness and horizon problem of the Hot Big Bang. The small quantum fluctuations predict spectra of perturbations, that are in accord with the observed cosmic microwave background temperature and polarization anisotropies (CMB) \cite{Planck:2018jri, Planck:2018vyg}.
  
However, this picture is not valid everywhere. In most models of Inflation, there are regions of parameter space where the quantum fluctuations are as large or even dominate the classical evolution.
If Inflation ever occurred in these regions, then we are in the so-called "Eternal Inflation" regime. In this regime universes are constantly being created due to quantum fluctuations with different CCs and possibly different physical laws in a never ending process, though each specific universe ends its inflationary phase at some point, giving rise to the so called "Multiverse". Given that any universe can arise with any physical laws tarnishes the predictability of Inflation as the predictions of Inflation now depend on some probability measure imposed on the Multiverse \cite{Ijjas:2013vea,Guth:2013sya,Ijjas:2014nta}.

For brevity, let us consider that from now on Inflation is described by a scalar field or fields with a potential. 
For a given model the region of the inflationary potential at the Eternal Inflation regime is not the region of the potential generating the observed CMB spectrum. The spectrum has to change from $P_s\sim 10^{-9}$ on CMB scales to $P_s> 1$ in the Eternal Inflation regime \cite{Goncharov:1987ir, Guth:2007ng}. Hence, Eternal Inflation requires some extrapolation, which can be avoided by suitable model building leaving all CMB predictions in tact \cite{2009arXiv0910.5515B,2010JCAP...09..007B,2014JCAP...03..054B,2020Symm...12..806A}.
Nevertheless, it is fair to say that such additional modeling is not warranted by any CMB observations, and without additional modeling, an Eternal Inflation regime and the ensuing Multiverse is rather generic.
In this note our assumption is that the Multiverse exists, Eternal Inflation is happening right now in some place, so the number of universes is not constant, and that it should abide by the laws of statistical physics and thermodynamics. If so, can we make predictions and test them?

Another troubling issue in contemporary theoretical physics is the smallness of the CC. Suggested resolutions include dynamical mechanisms \cite{2016JCAP...05..002B,Wetterich:1987fm,Bousso:2000xa,2021PhRvD.103L1303A}, statistical treatment \cite{Coleman:1988tj,Polchinski:1988ua} and many other suggestions, see various reviews \cite{1989RvMP...61....1W,Nobbenhuis:2004wn, Martin:2012bt}. No solution has been compelling so far. The Multiverse is the natural arena to try and predict the CC statistically, and various rather radical attempts even include anthropic considerations, for example \cite{Bousso:2007gp}.

We consider the Multiverse as an ensemble of universes with various CCs and volumes, and use standard statistical physics treatment.
This picture should be valid as long as we do not reach planckian temperature or energy density in various universes that require quantum gravity.
We show that the most likely or the average CC is exponentially suppressed, i.e. small and positive without resorting to anthropic arguments, solving the CC problem qualitatively. The result requires a "temperature" of the Multiverse that we define and is constrained to be subplanckian.

We then quantize the CC \'{a} la Sommerfeld, \cite{Xiang:2004pf}. The quantization allows us to sum the infinite series to fit the result to the observed CC. This in turn predicts the temperature of the Multiverse, which is near the GUT scale and suits the expectations of Inflation being near the GUT scale and the onset of Eternal Inflation to be in that regime.
Finally, the quantization allows us to view the system as a single Universe. This single Universe is a superposition of multiple different eigenstates with energy levels, rather than the physical existence of many universes.
Hence, the Multiverse by itself does allow to make non-trivial predictions without the need of a measure or anthropics, leaving Inflation with other problems such as the Big Bang singularity \cite{Borde:2001nh,Ben-Dayan:2016ydt,2016JCAP...09..017B,2019JCAP...07..050B,2020JCAP...05..015A,2020JCAP...12E.001B,2021JCAP...06..010A} , gradual ruling out of models and obviously the warm pool of quantum gravity conjectures \cite{Garg:2018reu,Ooguri:2018wrx,2019PhRvD..99j1301B,Akrami:2018ylq,2019JCAP...05..042A}.

We start by specifying our assumptions. 
We then define our thermodynamical system and show that the Multiverse predicts a small and positive CC in section \ref{sec:smallCC}. In section   
\ref{sec:quantization} we quantize the CC and discuss its implications. We then conclude.
\section{Assumptions regarding the Multiverse}
\begin{enumerate}
\item The Multiverse, if it exists, is described by statistical physics, where each Universe is a "particle" in the ensemble. As such the multiverse is a "bath" coupled to the system that is our universe or any other subsystem. These universes are semi-classical, i.e. their size is larger than Planck size\footnote{Actually, most of them should be huge or infinite. According to the standard reasoning, if eternal inflation ever occurred, then the overwhelming volume of the Multiverse should be dominated by universes with the largest possible CC.}.
\item The whole Multiverse system obeys the second law of thermodynamics and maximizes its entropy. 
\item The volume of the multiverse is the union of the volumes of all universes. For practical purposes, we will use the sum of volumes. 
\item The energy of the universe is the union of the energies in all the universes. For practical purposes, we shall use the sum of energies. 
\item We further simplify the analysis and consider that the energy density of each universe is captured by its CC and neglect possible additional energy contributions. We also consider only non-negative CCs. The rationale being that universes with negative CCs annihilate immediately and will not contribute to the most likely CC.

\item The semi-classical picture should hold as long as the energy density of each universe $\rho$ is lower than $M_{pl}^4$, where $M_{pl}$ is the reduced planck mass. At this point, quantum gravity should kick in and the analysis should break down. At energy density $M_{pl}^4>\rho>M_{pl}^2 V'^{2/3}$, where $V$ is the inflaton potential, we are in the Eternal Inflation regime, where universes are being produced copiously, but the semi-classical picture is still valid. The lower bound of this regime is our chemical potential $\mu$, see below. For $\rho<M_{pl}^2 V'^{2/3}$, we are not in the eternal inflation regime and universes are not produced. We will assume that the inflaton was high enough on the potential such that the Multiverse exists.

\item Since universes are being created and annihilated, we cannot use $N$ the number of universes to characterize the system. Rather, we use the chemical potential - how much does it cost to add a universe. 
For now, we assume that $\mu$ is the parameter that determines whether we are in the eternal inflation regime or not. When we are above this threshold, then universes are constantly being created, while when we are below, then they are not. 
In single field inflation, it corresponds to the value of the inflaton field where the scalar power spectrum $P_s>1$,  \cite{Goncharov:1987ir, Guth:2007ng}. This is because $P_s\sim\frac{V^{3/2}}{M_{pl}^3V'}$ up to some numerical factors. Depending on the realized inflationary model, this value of the field varies. For flatter potentials $\mu$ is lower. It is $\phi\sim 10^3 M_{pl}$ for the $1/2m^2\phi^2$ model, and $\phi\sim 20 M_{pl}$ for the Starobinski model. 
\item Since the number of universes and the overall total energy is not fixed, it makes sense to use the grand potential $\Omega(V,T,\mu)$, where $V$ is the total volume of the Multiverse and $T$ is its temperature that will be defined. If that temperature is larger than the planck mass, then again, quantum gravity analysis is required. Hence, our analysis is valid only for subplanckian temperatures $T\leq M_{pl}$. 
\end{enumerate}
\subsection{Relevant comments}
\begin{enumerate}
    \item The sole significance of the chemical potential is whether it is smaller than the typical energy (not energy density) of each universe. If so, then our analysis is valid, and the chemical potential can actually be neglected altogether.
    \item We are agnostic on how the universes should be classified as bosons, fermions or Boltzmann particles. Fermions- as is evident, no two universes are in the same quantum state, bosons - as there is no reason to require an antisymmetric wave function, Boltzmann particles - as we are dealing with very large systems that should behave classically.
    The main point is that in all cases we get the same result - a small CC.
    \item A reasonable modification of the assumption regarding the volume is that the union of volumes of the universes is a lower bound on the total volume of the Multiverse, i.e. $V=\sum_s V_s+\cal{V}$, where $V_s$ is the volume of the s-th universe. This will not alter the results of our derivation.
\end{enumerate}

Given all the above, let us try to derive a small CC in the Multiverse without resorting to anthropics.

\section{Model of the Multiverse}
\label{sec:smallCC}
We start with the most basic toy model one can think of. Each universe just has a different CC, $\Lambda_s$, and no other source of energy. As a result, the pressure of each universe is $p_s=-\rho_s=-\Lambda_s$, and also $\epsilon_s=\Lambda_s V_s$ is the energy of each universe, where $V_s$ is the volume of the universe. 

\subsection{Thermodynamical equilibrium}
Let us consider micro-canonical ensemble of universes with a given
$N$ total number of universes and $V$ total volume. We take $n_{sl}$ to be
the number of universes with volume  $V_s$ and a CC $\Lambda_l$ (where $s,l$ can be smaller/equal to $N$).
The energy and the entropy of this micro-canonical system is
\be
E=\sum _{s,l=1}^N n_{sl}  V_s \Lambda_l,
\ee
\be
S=N\ln N- \sum _{s,l=1}^N n_{sl}\ln {n_{sl}}.
\ee

 In order to obtain a thermodynamical equilibrium, we need a mechanism which changes the energy and the entropy of our micro-canonical ensemble, while keeping the total number $N$ and volume $V$ fixed. In order to do that, consider that a universe can (somehow) change its $\Lambda_l$ while keeping its volume $V_s$ unchanged. 
Whenever a universe with  $V_s$ and  $\Lambda_l$ changes to a universe with the same volume but with $\Lambda_l'$ the number of universes  $n_{sl}$ is decreased by one, whereas the number of universes $n_{sl'}$ is increased by one. Thus the energy of the ensemble changes by 
\be
\Delta E=(\Lambda_l' -\Lambda_l) V_s
\ee
and the entropy by
\be
\Delta S=S_f-S_i=
\ee
\be-(n_{sl'}+1)\ln(n_{sl'}+1)-(n_{sl}-1)\ln(n_{sl}-1)+n_{sl'}\ln_{sl'}+n_{sl}\ln_{sl}=
\ee
\be
=-\ln (n_{sl'}/n_{sl}).
\ee
 As a result, in equilibrium the temperature of the ensemble is 
\be 
\frac{1}{\beta}=T=\left(\frac{\Delta E}{\Delta S}\right)_{N,V}=-\frac{(\Lambda_l' -\Lambda_l) V_s}{\ln (n_{sl'}/n_{sl})}
\ee
where we took the Boltzmann constant to be unity.
Thus we have obtained the expected connection 
\be
\frac{n_{sl'}}{n_{sl}}=e^{-\beta(\Lambda_l' -\Lambda_l) V_s}
\ee
which leads to the probability density
\be
\rho_{sl}=A e^{-\beta \Lambda_l V_s}.
\ee

The above result gives us a canonical ensemble of universes with Boltzmann-like behavior. From now on each universe with $\Lambda_l$ and $V_s$ will be denoted by a single parameter $s$. We can then easily normalize the probability and get the standard expected partition function $Z=\sum_s e^{-\beta \epsilon_s}=\sum_s e^{-\beta \Lambda_s V_s}$. 
An immediate calculation of the average CC will yield:
\be \label{eq:LBoltzmann}
<\Lambda>=\sum_s\frac{\Lambda_s e^{-\beta\epsilon_s}}{\sum_i e^{-\beta\epsilon_i}}.
\ee
This allows the immediate sanity check that assuming all CCs are equal $\Lambda_s\equiv \Lambda$ one gets the expected result of $<\Lambda>=\Lambda$. Moreover, it is clear where the suppression of the CC may come from. As long as $\beta \epsilon_s \gg 1$, which is to be expected from large universes and low temperature, the contribution to the CC is expected to be exponentially small.  We can define a chemical potential in a similar fashion, which will lead us to a grand canonical ensemble, as the Multiverse should be. 

\subsection{The grand potential}

The Multiverse does not have a conserved number of universes, quite the opposite, we therefore define the grand potential with a chemical potential instead of the number of universes as a state variable. Deriving the grand potential proceeds in the same manner as described in the previous section for the canonical ensemble.
Consider the grand potential:
\be
\Omega(T,V,\mu)=-\frac{1}{\beta a}\sum_s \ln\left(1+a e^{-\beta(\epsilon_s-\mu)}\right).
\ee
Recall that $a=1$ for fermions, $a=-1$ for bosons, and the $a=0$ limit for classical Boltzmann indistinguishable particles. Using our assumptions the total volume and the total energy of the Multiverse are given by
\be
V=\sum_s V_s,\quad
U=\sum_s \epsilon_s=\sum_s \Lambda_s V_s.
\ee
i.e. the total volume of the multiverse is the sum of volumes of the universes and the total energy is the sum of energies. Let us note that there can be many universes with the same energy, CC or volume, the point is that we order them and take all of them into account by summing over $s$.
 Therefore, one can calculate the average pressure which will be the most likely pressure one measures by differentiating with respect to the volume $V$ while keeping $T,\mu$ fixed.
 Since $\Lambda=-p$, minus the average pressure is actually the most likely CC one expects to observe.
 \begin{multline}
<\Lambda>=-<p>=
\left(\frac{\partial \Omega}{\partial V}\right)_{T,\mu}=
\sum_s\frac{\partial \Omega}{\partial V_s}\frac{\partial V_s}{\partial V}= \\
-\frac{1}{\beta a}\sum_s \frac{-\beta a z \Lambda_s e^{-\beta\epsilon_s}}{\left(1+a z e^{-\beta\epsilon_s}\right)} = 
\sum_s \frac{\Lambda_s e^{-\beta(\epsilon_s-\mu)}}{\left(1+a e^{-\beta(\epsilon_s-\mu)}\right)}
\end{multline}

Let us contemplate the various possibilities and implications of this result. The crucial issue is the argument of the exponent. If $\beta(\epsilon_s-\mu) \gg 1$ then we get the fascinating result that the CC is a sum of exponentially suppressed contirbutions. As such it is parameterically suppressed compared to the naive expectation, of unweighted average and explains a \textit{a small and positive CC based on statistical physics arguments without any anthropics.}
As stressed, this is irrespective of whether the universes are bosons, fermions or Boltzmann particles as in all cases the denominator will practically be unity.\footnote{For bosons, $a=-1$ and then $0\leq e^{\beta \mu} \leq 1$, hence we know that the exponent is really decaying. For Boltzmann particles, $a=0$ and the denominator is unity by definition. Finally, regarding fermions, $a=-1$, as long as the chemical potential is $\mu \lesssim \mathcal{O} (10) M_{pl}$ as with currently allowed models, the argument of the exponent is still negative and therefore exponentially suppressed. Therefore, the chemical potential is rather unimportant for the validity of the procedure and can be neglected. Any large universe and/or "cold" Multiverse will be weighted with extreme exponential suppression $e^{-\beta \varepsilon_s}=e^{-\beta \Lambda_s V_s}$.} 
\be \label{eq:classical}
<\Lambda>\simeq \sum_s \Lambda_s e^{ -\beta (\epsilon_s-\mu)}=\sum_s \Lambda_s e^{ -\beta (\Lambda_sV_s-\mu)}
\ee


 
 The main question is then whether \eqref{eq:classical} is really the case. Minkowski universes have a vanishing contribution. Obviously infinite universes will not contribute to the CC as well. However, these infinite universes mean infinite energy, and we would like to regulate it. One way is that only closed universes contribute to the CC, while flat and open universes with infinite volume do not.
Alternatively, let us "weigh" all universes by energy within their dS horizon. The idea of a static dS patch is a recurring theme in discussions of the Multiverse, specifically in the context of the holographic principle and thermodynamics, for example \cite{Ben-Dayan:2020pbg,Dyson:2002pf,Susskind:2021omt,Susskind:2021yvs,Susskind:2021dfc}.
Another virtue, is the fact that for pure dS, the volume enclosed within the horizon is time-independent.
For a given dS space with a specific CC, $\Lambda_s$, we have:
\bea \label{eq:dshorizon}
V_s=4\pi\sqrt{3}\frac{M_{pl}^3}{\Lambda_s^{3/2}}, \quad &\Rightarrow& \quad \epsilon_s=\Lambda_s V_s= 4\pi\sqrt{3}\frac{M_{pl}^3}{\sqrt{\Lambda_s}}=  4\pi\sqrt{3}\frac{M_{pl}}{\alpha_s^2}\gg M_{pl}\cr
&\Rightarrow& <\Lambda>\simeq\sum_s \alpha_s^4 M_{pl}^4 e^{-\beta 4\pi \sqrt{3} M_{pl}/\alpha_s^2}
\eea
where we have used $\Lambda_s=\alpha_s^4 M_{pl}^4$, and in our conventions the dS horizon is at $R_c=\sqrt{3/\Lambda}M_{pl}$.
If $\alpha_s>1$ the energy density is super-planckian where the classical geometry of universes is not valid. So we must limit ourselves to $\alpha_s \leq 1$. 
Notice that using the dS horizon as a regulator, universes with planckian energy density are the least suppressed and will contribute the largest contributions to the average CC.
Also notice that all the terms in \eqref{eq:classical} are positive definite, which is worrisome. If one of them is larger than the observed CC, the idea fails. 
Consider a universe with planck energy density $\Lambda=M_{pl}^4$. Such a universe will contribute to the average CC,
\be
<\Lambda > \supset M_{pl}^4e^{-\beta 4\pi \sqrt{3}M_{pl}}<
\Lambda_{obs.}\simeq 10^{-119}M_{pl}^4
\Rightarrow T<\frac{4\pi\sqrt{3}}{119 \ln 10}M_{pl} \simeq 0.4 M_{pl}
\ee
As long as the temperature of the Multiverse is subplanckian, as we require in our semi-classical treatment, even universes with planckian energy density will not destroy the small CC.
More generally, assuming some distribution of CCs in the different universes, the observed CC in our universe will constrain the temperature of the Multiverse.
Stepping into the mine field of distribution of CCs, let us assume that $0\leq \alpha_s \leq 1$ and change the sum to an integral over $\alpha_s$, yielding a continuum of CCs and universes. The integral converges nicely:
\bea
<\Lambda >&=&M_{pl}^4\int_{0}^{1}d\alpha_s \alpha_s^4 e^{-\beta 4\pi \sqrt{3} M_{pl}/\alpha_s^2}\cr
&=&
M_{pl}^4\times\frac{1}{15}e^{-b}\left(3-2b+4b^2-4\sqrt{\pi}b^{5/2} e^{b}\, \text{erfc}[\sqrt{b}]\right),\quad b\equiv \beta 4\pi \sqrt{3} M_{pl} \\
\Rightarrow T&\simeq& 0.08 M_{pl},\quad  <\Lambda>=\Lambda_{obs.}
\eea

Hence, even if we allow a continuum of universes and CCs, at least for the simplest distribution the integral converges, with a valid temperature of the Multiverse. Hence, this method is applicable not just for a countable number of universes, but also for a continuum distribution. Given the exponential suppression and the fact that $0\leq \alpha_s \leq 1$, many other measures $\int_{0}^{1}d\alpha_s f(\alpha_s) \alpha_s^4 e^{-\beta 4\pi \sqrt{3} M_{pl}/\alpha_s^2} $ will still converge.


To summarize, we have shown that treating the Multiverse as an ensemble of universes dominated by their cosmological constants, yields a prediction of an exponentially small positive cosmological constant without using anthropic arguments. In this sense the Multiverse is predictive. We have used standard statistical physics analysis and did not require some special measure of probability. The only constraint is on the temperature of the Multiverse $T<M_{pl}$ which is needed anyway for self-consistency. Considering a flat measure yielded the observed CC with the Multiverse temperature being $T\simeq 0.08 M_{pl}$. To substantiate this result, we shall take an additional step assuming the CC is quantized \'{a} la Sommerfeld. The quantization will allow a more precise determination of the CC.


   \section{Quantization of the CC}
  \label{sec:quantization}
  We have shown that the Multiverse qualitatively predicts a small and positive CC, with rather mild constraints. However, except knowing that the temperature is subplanckian we know very little about the would be thermodynamics of the Multiverse. Assuming a distribution of CCs, one can derive the temperature of the Multiverse given the observed CC as we have done. Unfortunately, we also know very little about the would be distribution of CCs.
  We therefore add another assumption which is the quantization of the CCs.
  Let us quantize the area of the dS horizon to be \cite{Xiang:2004pf}:
  \bea
  \mathcal{A}_s&=&8\pi\left(n_s+\frac{1}{2}\right) \chi\, l_{pl}^2=4\pi\left(2n_s+1\right) \chi\, l_{pl}^2=\frac{4\pi\times 3 M_{pl}^2}{\Lambda_s}\\
  \Rightarrow \Lambda_s&=& \frac{24 \pi M_{pl}^4}{(2n_s+1)\chi}
  \eea
  where $\chi$ is some $\mathcal{O}(1)$ number and $n_s$ is an integer for each universe, and the numerical factors are due to the fact that we have the Planck length and the reduced Planck mass. For large universes $n_s$ is expected to be huge.
  Substituting the quantized CC into our formula for the average CC using the horizon volume \eqref{eq:dshorizon} gives, 
  \be 
  \label{eq:Lquantized}
  <\Lambda>=\sum_{\{n_s\}}\frac{\frac{24 \pi M_{pl}^4}{(2n_s+1)\chi}e^{-\beta \sqrt{2\pi\chi(2n_s+1)} M_{pl}}}{1+a e^{-\beta\sqrt{2\pi\chi(2n_s+1)} M_{pl}}}
  \ee
where we need to sum over the possible $\{n_s\}$, i.e. how many universes have the same CC and therefore the same $n_s$. Again, in principle we can write \eqref{eq:Lquantized} with the chemical potential but again it will be negligible. Given the huge number of universes and the fact that $n_s$ could be a huge number, we replace the complicated sum by simply a sum over $n$ to infinity.
\be \label{eq:Lsummed}
  <\Lambda>=\sum_{n_s}\frac{\frac{24 \pi M_{pl}^4}{(2n_s+1)\chi}e^{-\beta \sqrt{2\pi\chi(2n_s+1)} M_{pl}}}{1+a e^{-\beta\sqrt{2\pi\chi(2n_s+1)} M_{pl}}}.
  \ee

A rather crude but good approximation is simply taking into account the lowest order term, $n_s=0$:\footnote{Better approximations using Bernoulli numbers will not change the result in a meaningful manner.}
\be
<\Lambda>\simeq \frac{24 \pi}{\chi}M_{pl}^4 e^{-\beta\sqrt{2\pi\chi} M_{pl}}+\cdots
\ee
where we have used the fact that the argument in the exponent is large, which explains the domination of the first term as well as the fact that the denominator is approximately unity.
Hence, we have achieved an exponential suppression of the CC as well as a quantitative estimate of it.
Taking $<\Lambda>=\Lambda_{obs.}=10^{-119} M_{pl}^4$ as the observed value of the CC, we can derive the temperature of the Multiverse:
\be
T=-\frac{\sqrt{2\pi \chi}}{\ln(\chi/24 \pi)+\ln \left(\frac{<\Lambda>}{M_{pl}^4}\right)}M_{pl}\sim 10^{-2} M_{pl}.
\ee
Hence, according to our analysis the temperature of the Multiverse is around the GUT scale which is the natural scale of Inflation. This is also roughly the scale where $V_{inf}^{1/4}$ is in the Eternal Inflation regime, providing us with two reasonable consistency checks.

\section{Discussion}
We have assumed that the Multiverse exists as a semi-classical entity due to Eternal Inflation, and applied the standard statistical physics techniques to predict the CC. As such, the Multiverse has thermodynamical quantities such as temperature, pressure etc.,  and we have defined these quantities.

Using the fact that the CC is also the negative pressure of the would be Multiverse, we were able to derive the average CC. For large systems such as the Multiverse, such average is practically a certainty, so this average should be the observed CC in an overwhelming number of universes in the Multiverse and therefore in ours as well.

We have shown that the average is a sum of exponentially suppressed contributions, and therefore is expected to be exponentially small.
Using the dS horizon for evaluating the volume of universes, we got that the temperature of the Multiverse is only constrained to be less than the planck scale, which is required anyway to avoid quantum gravity effects. Hence, in the Multiverse the CC should be exponentially small and positive without the need for some anthropic considerations.

Considering the simplest possible continuum limit, we derived the temperature of the Multiverse to be $T\sim 0.08 M_{pl}$. It is also clear that this continuum limit is converging for a large class of possible CC distributions. So our analysis is applicable even if the process of universe generation, and the allowed value of CC is continuous.
For a more rigorous derivation of the CC we quantized the area of the dS horizon. The result gave a prediction for the temperature of the Multiverse to be $T\sim 10^{-2} M_{pl}$. This should be the expected temperature a-priori since the natural scale of inflation in accord with observations is the same scale, and it also suits the scale of the Eternal Inflation regime. We therefore view this prediction of the temperature as two consistency checks of our suggestion. 
More broadly, we have traded the small value of the CC for a Multiverse grand canonical ensemble with temperature $T\sim 0.01-0.1 M_{pl}$. Support for our prediction can come either from an independent derivation of the temperature of the Multiverse from other considerations, or a measurement that can be interpreted as measuring the temperature. From our knowledge about dS temperature and dimensional analysis we can speculate that the Multiverse temperature will be $T\sim M_{pl}/(2\pi \sqrt{3})$, which results in a CC of $<\Lambda>\sim M_{pl}^4 e^{-24 \pi^2 \mathcal{O}(1)}\sim 10^{-103 \times \mathcal{O}(1)} M_{pl}^4$. Nevertheless, we expect that the temperature of the Multiverse will be determined differently from the temperature of a single universe and one should not confuse between them. 

We would like to make two additional remarks.

We have analyzed universes with a dominant CC but there may be additional energy components, such as in our very own Universe. To that aim, our method can be extended. Taking into account universes with pressureless dust and a CC as well will not qualitatively modify our result. The energy in the exponent of such a term will be larger due to the dust contribution, but it will not modify the average pressure and therefore the average CC. Hence, our method is not limited to empty universes. 
Moreover, addressing the empty universe problem \cite{Bousso:2007gp}, we envisage the following:
During eternal inflation universes are created with different and random volumes and CCs. These universes equilibrate and then decouple from one another. At this very short stage they get to thermal equilibrium by changing their original CC. Then, each universe evolves on its own. i.e. it can inflate, reheat, etc.

Finally, as explained in the introduction, it could be that the Multiverse does not exist, as it requires extrapolation of the inflationary potential way beyond the region of CMB observations. The quantization of the CC allows us to interpret the Multiverse in a different manner. The above system \eqref{eq:Lsummed} is not necessarily an average over a multitude of universes with different physical laws, but rather a thermal average of a single Universe with quantized energy levels. However, in that case 
one needs to rethink how a thermalization process is generated, for example by tracing over degrees of freedom. As is well known from statistical physics, tracing over one of two systems is enough for the untraced system to behave as a thermal system. This opens up interpreting our Universe as coupled to other universes in other ways such as Braneworld scenarios, globally or locally interacting, for example \cite{Coleman:1988tj,Polchinski:1988ua,1989RvMP...61....1W,Nobbenhuis:2004wn, Martin:2012bt,Bondarenko:2020gkk}.
It would be interesting to try and work out such a model, and test our analysis.

\textbf{Acknowledgements:}
We thank Marcelo Schiffer for useful discussions and Michal Artymowski for discussions on the volume of dS space.
\bibliography{main}
\bibliographystyle{JHEP}

\end{document}